# Evolutionary perspective of large language models on shaping research insights into healthcare disparities

David An

## Abstract

**Introduction**. Advances in large language models (LLMs) offer a chance to act as scientific assistants, helping people grasp complex research areas. This study examines how LLMs evolve in healthcare disparities research, with attention to public access to relevant information.

**Methods**. We studied three well-known LLMs—ChatGPT, Copilot, and Gemini. Each week, we asked them a consistent prompt about research themes in healthcare disparities and tracked how their answers changed over a one-month period.

**Analysis**. The themes produced by the LLMs were categorized and cross-checked against H-index values from the Web of Science to verify relevance. This dual approach shows how LLMs' outputs develop over time and how such progress could help researchers navigate trends.

**Results**. The outputs aligned with actual scientific impact and trends in the field, indicating that LLMs can help people understand the healthcare disparities landscape. Time-series comparisons showed differences among the models in how broadly and deeply they identified and classified themes.

**Conclusion**. The study offers a framework that uses the evolution of multiple LLMs to illuminate AI tools for studying healthcare disparities, informing future research and public engagement strategies.

## Introduction

Artificial intelligence, particularly large language models (LLMs), have shown their capacity for high-quality text comprehension and generation across a diverse array of topics (Chang, 2024). These models, exemplified by ChatGPT, have sparked considerable academic and public interest due to their accessibility and generative capabilities (Wang, 2024). Their rapid development in generative ability offers a unique avenue to scientific inquiry into various complex domains like healthcare research, a field increasingly embracing AI to improve clinical documentation, patient engagement, professional medical knowledge, and research discovery (Wang, 2024; Nazi, 2024;Wang, 2024; Wang, 2024; Schlicht, 2025; Busch, 2025; Maity, 2025; Singh, 2023; Singhal, 2023; Yu, 2025).

Within this field, healthcare disparities represent a specific sector posing critical global challenges, characterized by inequitable differences in health outcomes and access to care between demographic groups that arise from socioeconomic, environmental, historical, cultural, and systemic factors (Fullman, 2018; Ong, 2024). These disparities can be further exacerbated by unexpected events like the COVID-19 pandemic, which underscores the need for innovative tools to quickly and accurately navigate the research landscape in this field, thereby enabling timely discovery of any unmet needs (Ong, 2024).

The progressive adaptability in LLMs potentially translates to unprecedented opportunities to explore how the research landscape of healthcare disparities has been evolving (Meduri, 2024; Maity, 2025). Not only can LLMs synthesize vast scientific literature, but they can also help trace shifts in research focus over time — revealing emerging themes, unmet needs, and gaps in healthcare equity (Meduri, 2024; Yu, 2025). By tracing LLM-generated insights, it is possible to empower both the public and professionals to swiftly update their knowledge on the shifting priorities, evolving methodologies, and changing societal contexts with respect to the research effort in mitigation of healthcare disparities (Ray, 2023; Ong, 2024).

However, concerns are raised about potential discrepancies in the insights derived from using and relying on LLM technology. These include biases embedded in training data, variable accuracy across different domains, inconsistencies in multilingual outputs, and lack of domain-specific optimization: all of which may compromise the integrity and fairness of LLM-derived insights (Nazer, 2023; Yang, 2023; Clusmann, 2023; Beheshti, 2025; Schlicht, 2025).

Biased, inconsistent, and unreliable responses over time from LLMs across models might exacerbate knowledge deficiencies in healthcare disparities as the general public increasingly relies on AI chat bots to obtain information, which is expected given their user-friendly interfaces and relative accessibility (Thirunavukarasu, 2023; Veinot, 2018). The potential for LLMs to inadvertently perpetuate biases or spread misinformation underlines the urgency of addressing these limitations through critical oversight and robust governance (Maity, 2025; Wang, 2024).

The project herein explores this intricate concern by examining the adaptive behaviour of LLMs when applied to healthcare disparities research from the public's perspective. Specifically, the study is designed to discover thematic foci, identify gaps, and assess the reliability and biases in LLM outputs by comparative evaluation of information generated from various models regarding healthcare disparities research. Integrating these evolutionary responses from LLMs over time would be particularly beneficial for researchers and policymakers to navigate intricacies of the healthcare disparities field in a comparative manner, allowing them to make better informed decisions about what areas require greater attention.

## Methods

The LLMs included in the study were ChatGPT from OpenAI (chat.openai.com), Gemini from Google (gemini.google.com), and Copilot from Microsoft (copilot.microsoft.com) (Ray, 2023; Google, 2023; Mehdi, 2023). This inclusion was based on consideration of several key factors: their significant market dominance and broad user accessibility as major AI developers, diverse underlying architectures, demonstrated general-purpose utility, and continuous development and updates. This selection would reflect the current landscape of accessible and widely adopted artificial intelligence tools, thereby enhancing the reproducibility and interpretability of the research for a broad audience.

The same prompt was executed in the chat box of the LLMs to evaluate how technology could assist with shaping comprehension of a specific area of research. The field of healthcare disparities research was used as an illustrative case in this study given its present relevance and developing nature of complexity (Braveman, 2006; Riley, 2012). The prompt given to each LLM was:

> 'Explore and identify 20 top research topics related to healthcare disparities across various scientific databases (Web of Science, PubMed, and Scopus). Each topic should be listed in less than five words and categorized as either an established or emerging theme'.

The query was meant to elicit primary research themes that AI would recognize as the most significant in healthcare disparities, with an idea of categorization into newly emerging or historically established. The method of data acquisition ensures that the public can easily access and obtain pertinent information. The query was initially executed in the LLMs on March 25, 2024, and then repeated weekly each Monday for five consecutive weeks to observe how responses would evolve over time in terms of theme identification and classification into either established or emerging.

For a theme that transitioned between established and emerging in a LLM's responses across the study period, determination of its final classification would be based on the frequency of being classified as one of the statuses. For example, if there were more classifications of a theme being established, it's considered established. Conversely, for a theme receiving an equal number of emerging and established classifications, it would depend on the most recent identification.

The themes recognised by the LLMs were validated through the indexed record of research publications using the Citation Report analysis tool carried by the Web of Science database. For this, the search term 'ALL = (healthcare disparities) AND ALL = (the topic of interest)' was used to retrieve the relevant publication record for each identified research theme.

Subsequently, the Citation Report analysis tool was applied on the publication record to examine the productivity and scientific impact of the research theme based on the H-index values. The H-index serves as a valuable metric, with a higher value indicating a more mature or influential research theme (Costas, 2007; Norris, 2010; Shah, 2023). By comparing the H-indices of the identified research themes, it directly facilitates the validation of the LLMs' efficacy in their identification and classification, allowing for an assessment of the themes' relative impact. For instance, a research theme with an H-index of 100, as opposed to 25, signifies that the associated publications within the former are collectively impactful, as each of the 100 papers have been cited at least 100 times.

If the LLMs repeatedly classify such a highly indexed theme as established, it strongly suggests successful identification and classification. In this manner, the simultaneous assessment across the three LLMs with additional metric verification by H-index values would allow for a parallel comparison and dual validation of the unique capabilities and limitations of different AI tools. For statistical significance, histograms were generated to visualize the distribution of H-indices across all the identified research themes, providing insights into the central tendency, spread, frequency, and overall shape of theme impact. Non-parametric Kruskal-Wallis H test was utilized to evaluate the performance of the LLMs in identification and classification of research themes across three metrics, H-index distributions, frequency of theme identification, and classification into established or emerging themes. The conventional alpha level of 0.05 was adopted as the significance criterion for all the statistical tests. This combination of descriptive visualization with inferential non-parametric testing offered a comprehensive and statistically sound framework for validating the quality and distinctiveness of research theme identification from the LLMs.

## Results and discussion

### General trend of LLM responses

This study followed ChatGPT, Copilot and Gemini on their analysis of the primary research themes in healthcare disparities over a five-week period. The outcomes represent a positive initial assessment of using LLMs from the perspective of the public. The data reveal the adaptive responses of the LLMs over time and their significant variation in shaping the research landscape of healthcare disparities, both in the breadth of themes identified and in the depth of categorizing them as either 'established' or 'emerging' (Table 1).

Partial research themes remained consistently relevant at a certain level, while others evolved or emerged in prominence. For themes which were consistently recognized, only 'mental health disparities' was generated by all the three LLMs in every week of the study, suggesting its high relevance as a major topic in healthcare disparities research (Cook, 2019, 2007; Miranda, 2008; Bailey, 2018). Other than mental health disparities, many other themes were present within an LLM's responses continually throughout the five weeks. Seven of these themes were consistently classified as 'established', suggesting continuous emphasis and recognition of their long-standing nature and ongoing importance, and the others repeatedly appeared as 'emerging', reflecting their recent and substantial attention in healthcare disparities research. The consistency of themes in the responses of an LLM indicates that these are well-realized issues and continue to receive extensive interest in research (FitzGerald, 2017; Fiscella, 2016; Berkman, 2010; Kirubarajan, 2021). In particular, consistent recognition of new focus themes reflects recent developments and shifts in the healthcare landscape, e.g., thematic recognition of climate change and health AI points to an emerging interest in integrating environmental and technological components to better understand and combat disparities (Nogueira, 2022; Abràmoff, 2023 ). Less consistently, there were themes identified intermittently across the weeks, such as immigrant health, cultural competency, and geographic disparities. Apart from these instances, the other themes were detected much less frequently by any of the three LLMs, implying that they are probably niche or specialized areas of research that are not as broadly covered in the field of healthcare disparities.

### H-index validation of LLM responses

Corroboration with the Web of Science H-index data provides additional metric verification for the LLM application in terms of its theme identification and classification. A majority of the themes recognized by the LLMs had H-index values from 3 to 119. The most frequent range fell between 3 and 32, followed by a cluster from 32 to 90, and a very small proportion exceeded 119 as illustrated by the histogram analysis (Figure. 1A). This distribution pattern acknowledges that the LLMs have the capability to identify a broad range of themes in healthcare disparities, signifying predominance for areas with moderate research activity rather than the highest impact. Themes recognized only once throughout the study period tended to have lower H-index values, typically below 50 (Table 1; Figure 1B). Overall, there was a clear distinction in the H-index values between established and emerging themes (Figure 1C). Themes categorized as established generally possessed an H-index above 50, reflecting their long-standing recognition and impact within the research community. On the other hand, emerging themes primarily fell below that threshold, indicating that they are at a nascent stage of development and have not yet accumulated substantial citations.

| | | | | | | | | | | | | |
|---|---|---|---|---|---|---|---|---|---|---|---|---|
| **ChatGPT** | Disparities in Cancer Care **(111)** | ■ | | | | | Infectious Diseases **(72)** | | ■ | ■ | | ■ |
| | Racial & Ethnic Health Disparities **(129)** | ▒ | ▒ | ■ | ■ | ■ | Digital Health Inequities **(13)** | | ▒ | | ▒ | |
| | Healthcare Access & Quality **(118)** | ■ | ■ | ■ | ■ | ■ | Genomics & Precision Medicine **(12)** | | ▒ | ▒ | ▒ | ▒ |
| | Socioeconomic Status & Health **(112)** | ■ | | | ■ | ■ | Health Literacy **(64)** | | ▒ | ▒ | ■ | ■ |
| | Nurse Bias & Patient Care Disparities **(16)** | ■ | | | | | LGBTQ+ Health Disparities **(24)** | | ▒ | ▒ | | |
| | Mental Health Disparities **(111)** | ■ | ■ | ■ | ■ | ■ | Disability & Healthcare Access **(47)** | | ▒ | ▒ | | |
| | COVID-19 & Healthcare Disparities **(66)** | ▒ | | | ▒ | | Veterans' Health **(109)** | | ▒ | ▒ | | |
| | Disparities in Disaster Healthcare **(18)** | ▒ | | | | | Immigrant & Refugee Health **(19)** | | ▒ | ▒ | ▒ | |
| | Pay-for-Performance & Disparities **(30)** | ▒ | | | | | Healthcare Policy Impacts **(79)** | | ■ | ■ | | |
| | Rural Healthcare **(75)** | ▒ | ▒ | ▒ | | | Geriatric Healthcare Access **(14)** | | ■ | ■ | | ■ |
| | Implicit Bias in Healthcare Providers **(40)** | ▒ | | | | | Health Technology & Inequality **(23)** | | | ▒ | | |
| | Health Equity Strategies **(42)** | ▒ | | | | | Gender Differences in Healthcare **(81)** | | | | ■ | ■ |
| | Barriers to Seeking Health Care **(44)** | ▒ | | | | | Impact of Insurance on Healthcare **(60)** | | | | ■ | |
| | Maternal & Child Health **(57)** | ▒ | ■ | ■ | | ■ | Healthcare Workforce Diversity **(25)** | | | | ■ | |
| | Cultural Competency **(55)** | ▒ | ▒ | ▒ | | | Geographic Disparities **(65)** | | | | ■ | |
| | Impact of Social Determinants **(78)** | ▒ | ■ | ■ | ▒ | ▒ | Climate Change **(15)** | | | | ▒ | ▒ |
| | Access to Mental Health Services **(71)** | ▒ | | | | | Intersectionality in Healthcare **(24)** | | | | ▒ | |
| | Racial Residential Segregation **(30)** | ▒ | | | | | Telemedicine & Access Issues **(38)** | | | | ▒ | ▒ |
| | Health Information & Communication **(16)** | ▒ | | | | | Health AI & Bias **(17)** | | | | ▒ | ▒ |
| | Ethics & Health Disparities **(34)** | ▒ | | | | | Health Data Privacy **(16)** | | | | | ▒ |
| | Environmental & Occupational Health **(16)** | | ▒ | ▒ | | | Global Health Initiatives **(21)** | | | | | ▒ |
| | Chronic Disease Management **(53)** | | ■ | ■ | ■ | ■ | Pandemic Preparedness **(17)** | | | | | ▒ |
| | Nutrition & Food Security **(11)** | | ■ | ■ | | | Technological Innovations **(11)** | | | | | ▒ |
| **Copilot** | Disparities in Cancer Care **(111)** | ■ | ■ | | | | Disparities in Aging Populations **(114)** | ▒ | ▒ | ▒ | ■ | ▒ |
| | Mental Health Disparities **(111)** | ■ | ▒ | ■ | ■ | ■ | Health Disparities in AI Ethics **(6)** | ▒ | | | | |
| | Healthcare Access **(118)** | ■ | | | | ■ | Racial Bias in Pain Management **(13)** | | ■ | | | |
| | Racial Disparities **(129)** | ■ | | ■ | | ■ | Maternal Mortality Disparities **(42)** | | ■ | ■ | ■ | |
| | Health Literacy **(64)** | ■ | ■ | ■ | ▒ | ■ | Access to Mental Health Services **(71)** | | ■ | | | |
| | Social Determinants **(78)** | ■ | ■ | ■ | ■ | ■ | Cultural Competency **(55)** | | ■ | ■ | ■ | ■ |
| | Health Equity **(80)** | ■ | | ■ | ▒ | ■ | Care Access for Undocumented Immigrants | | ▒ | | | |
| | Language Barriers **(58)** | ■ | ■ | ■ | ■ | ■ | Disparities in COVID-19 Vaccination Rates **(18)** | | ▒ | | | |
| | Obesity Disparities **(69)** | ■ | | | | | Health Disparities in Rural Areas **(75)** | | ▒ | | ■ | ■ |
| | Disability Disparities **(64)** | ■ | ▒ | ▒ | ■ | ▒ | Implicit Bias in Clinical Decision-Making **(11)** | | ▒ | | | |
| | LGBTQ+ Health Disparities **(24)** | ■ | ■ | ▒ | ▒ | ▒ | Access Barriers **(86)** | | | ■ | | |
| | Geographic Disparities **(65)** | ■ | | ■ | ■ | | Health Technology **(23)** | | | ▒ | ▒ | |
| | Digital Health Disparities **(13)** | ▒ | | ▒ | | ▒ | Immigrant Health **(54)** | | | ▒ | | ▒ |
| | Precision Medicine Disparities **(28)** | ▒ | ▒ | ▒ | ▒ | ▒ | Racial Bias in Diagnostics **(11)** | | | | ■ | |
| | Telehealth Equity **(11)** | ▒ | ▒ | ▒ | ▒ | ▒ | Access to Clinical Trials **(45)** | | | | ▒ | |
| | Environmental Justice in Health **(16)** | ▒ | ▒ | ▒ | ▒ | ▒ | Healthcare Workforce Diversity **(25)** | | | | ▒ | |
| | Genomic Health Disparities **(28)** | ▒ | ▒ | | | ▒ | Disparities in Indigenous Communities **(21)** | | | | ■ | |
| | Intersectionality in Healthcare **(24)** | ▒ | | ▒ | | ▒ | Maternal & Child Health **(57)** | | | | | ■ |
| **Gemini** | Race & Ethnic Disparities **(129)** | ■ | ■ | | | ■ | Structural Inequities **(32)** | ▒ | | | | |
| | Socioeconomic Status & Health **(112)** | ■ | ■ | | | ■ | Immigrant Integration **(9)** | ▒ | | | | |
| | Access Disparities **(144)** | ■ | | | | | Age & Technology **(47)** | ▒ | | | | |
| | Quality Disparities **(173)** | ■ | | | | | Maternal Health **(64)** | | ■ | ■ | ■ | ■ |
| | Chronic Disease **(91)** | ■ | | ■ | ■ | ■ | Language Barriers **(58)** | | ■ | ■ | ■ | ■ |
| | Mental Health Disparities **(111)** | ■ | ■ | ■ | ■ | ■ | Disparities in Cancer Care **(111)** | | ■ | ■ | ■ | |
| | Telehealth Access **(11)** | ■ | ▒ | ■ | ▒ | ▒ | Cardiovascular Risk **(82)** | | ■ | | | |
| | Implicit Bias **(40)** | ■ | ■ | ■ | ■ | ■ | Intersectionality in Healthcare **(24)** | | ▒ | ▒ | ▒ | |
| | Social Determinants **(78)** | ■ | | ■ | ■ | | Genomics & Disparities **(28)** | | ▒ | | | ▒ |
| | Rural Healthcare **(75)** | ■ | ■ | ■ | ■ | ■ | Stigma & Mental Health **(38)** | | ▒ | | | |
| | Digital Divide **(35)** | ▒ | ■ | | | ■ | Community-Based Interventions **(38)** | | ▒ | ▒ | ▒ | ▒ |
| | AI in Disparities **(48)** | ▒ | ▒ | ▒ | ▒ | ▒ | Big Data & Health Equity **(11)** | | | ▒ | | ▒ |
| | Climate Change **(15)** | ▒ | ▒ | ▒ | ▒ | ▒ | LGBTQ+ Health Disparities **(24)** | | | ▒ | | ▒ |
| | Social Media & Health Information **(25)** | ▒ | ▒ | ▒ | ▒ | ▒ | Cultural Competency **(55)** | | | ■ | ▒ | ■ |
| | Environmental Justice & Health **(16)** | ▒ | ▒ | ▒ | ▒ | ▒ | Stigma & Healthcare Access **(35)** | | | ▒ | ▒ | |
| | Precision Medicine & Equity **(28)** | ▒ | ▒ | ▒ | ▒ | | Health Insurance **(106)** | | | | ■ | |
| | Food Deserts **(3)** | ▒ | | | | | Geographic Disparities **(65)** | | | | ■ | ▒ |

**Table 1.** Healthcare disparities research themes identified by LLMs over a five-week period.

The query was executed weekly starting from 25th March, 2024. Bold numbers in parentheses denote the H-index value sourced from Web of Sciences for the corresponding theme. Black shading indicates an 'established' theme, and grey shading represents an 'emerging' one. Blank grids indicate themes not identified in that week.

The alignment between the theme classification and the metric measure provides scientific evidence in favour of using LLMs as a valuable tool for navigating the evolving landscape of healthcare disparities research. While opinions vary on the algorithmic fidelity in other domains (Amirova, 2024; Argyle, 2023), the results here support the idea that LLM outputs are a reliable reflection of underlying trends and academic significance in healthcare disparities, taking into account the level of research maturity, rather than being random or arbitrary.

Combining LLM outputs with relevant metric data analyses can lead to a more thorough and trustworthy interpretation of both established research and emerging frontiers in healthcare disparities. Practically, it can enhance fast identification of key research areas, inform funding decisions, support strategic research planning, optimize resource allocation to areas with the greatest potential for impact, and ultimately advance the impactful healthcare research.

For instance, having a clear awareness of theme classifications can help prioritize research efforts and funding - established themes with high H-index values often reflect areas with proven impact and ongoing relevance, and emerging themes with low H-index values but high identification frequency may represent burgeoning areas that could benefit from increased attention and resources. Nevertheless, further study is needed to solidify these findings and facilitate a more systematic incorporation of LLMs into the workflow of healthcare disparities research, thereby accelerating advancements in this critical field.

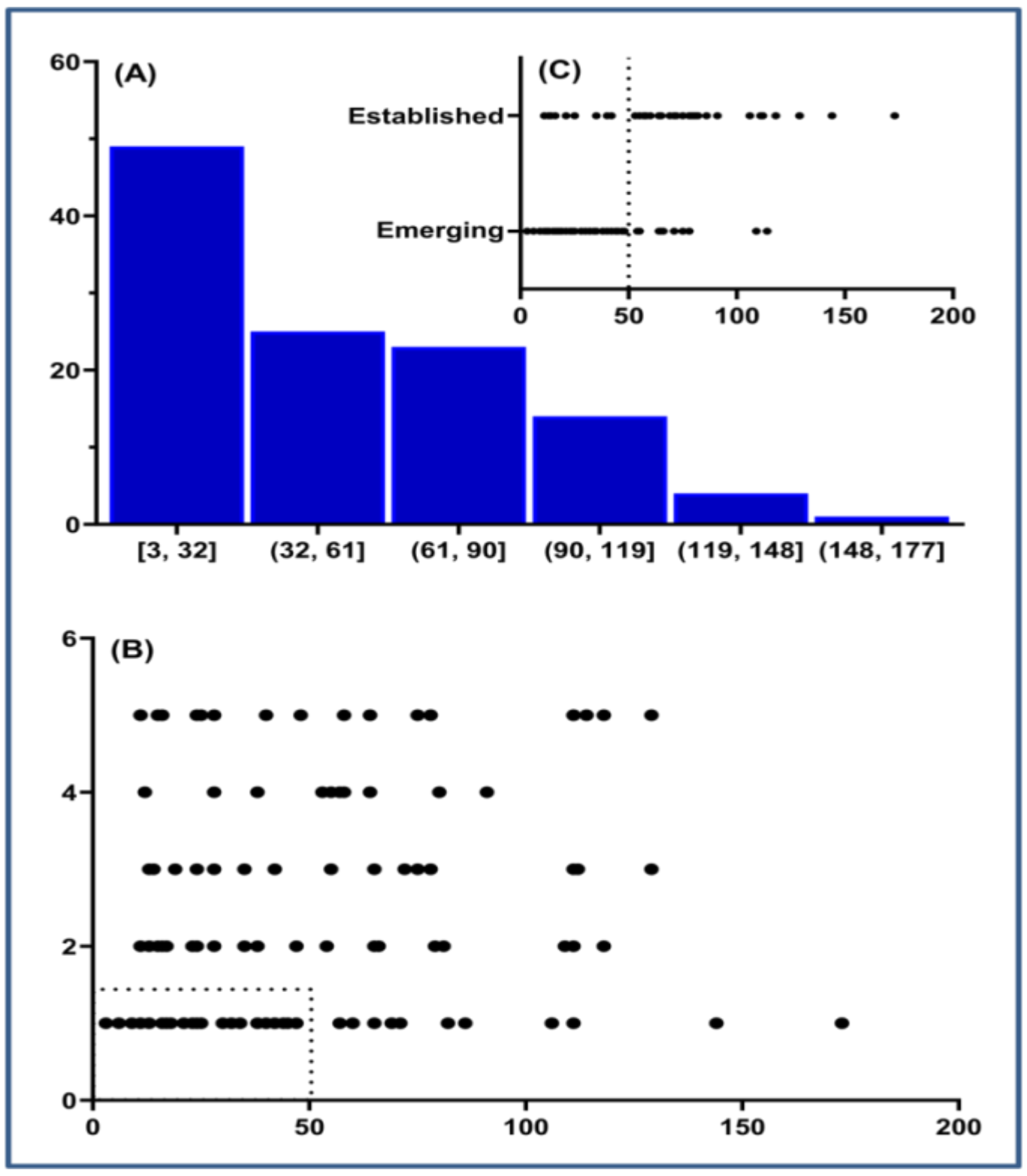

**Figure 1.** The relationship between healthcare disparities research themes identified by LLMs and their corresponding H-index in the Web of Science database. (A) Histogram showing distribution (y-axis) of the themes in the H-index range (x-axis). (B) Frequency (y-axis) of a theme identification corresponding to its H-index value (x-axis) throughout the five weeks. (C) Classification of an identified research theme (y-axis) corresponding to its H-index value (x-axis).

## Comparison of LLM-specific responses

LLM-generated responses have exhibited distinct thematic foci among the three models, diverging from a general trend. ChatGPT notably distinguished research themes related to racial and ethnic health disparities, healthcare access and quality, and social determinants; Copilot identified health literacy, language barriers, and disability disparities as prominent topics; Gemini recognized implicit bias, rural healthcare, and climate change to be important aspects of research on healthcare disparities.

Specifically, some themes were differentially identified only once in an LLM over the study period, like health data privacy and pandemic preparedness from ChatGPT in the 5$^{th}$ week, racial bias in pain management from Copilot in the 2$^{nd}$ week, and structural inequities from Gemini in the 1$^{st}$ week. The sporadic introduction of emerging themes in the responses may suggest an expansion of the LLMs' reach into prospective issues as a proactive approach to incorporating cutting-edge topics and societal issues (e.g., global health initiatives and technological innovations). Notably, there were themes transitioned between being classified as 'emerging' and 'established' over the study period (e.g., social determinants from ChatGPT, and digital divide and telehealth access from Gemini). In practical terms, it could suggest that these areas are highly complex and deeply integrated with emerging issues and multi-level fields of research regarding healthcare disparities (Brondolo, 2009; Ruprecht, 2021; Saeed, 2021).

The LLMs' evolving responses over a relatively short period reveal their ability to adapt to ongoing academic and societal changes, likely driven by updates in the training data or changes in underlying algorithms (Aldoseri, 2023).

When comparing the generative capabilities of the three LLMs, ChatGPT presented a highly dynamic approach and displayed significantly greater variability than Copilot and Gemini in responses to the queries of healthcare disparity research themes. This is evident from the data that over 50% of the themes from ChatGPT's responses varied from week to week over the course of the study (Figure 2A). This aligns with ChatGPT's ability to browse the web to continuously scrape data, capable of updating information in real-time (OpenAI, 2024).

There was an especially consistent shift in emerging themes, notably implicit bias in the 1$^{st}$ week, environmental health in the 2$^{nd}$ week, health technology in the 3$^{rd}$ week, intersectionality in the 4$^{th}$ week, and health data privacy in the 5$^{th}$ week. ChatGPT's greater propensity for week-to-week variation indicates its exceptional ability to capture new and evolving research themes, as well as a high tendency to discover extremely specific topic areas.

Relative to the established group of research areas, all the three LLMs produced more variable responses for emerging themes, with Copilot exhibiting the least variability and greatest consistency in recognizing both well-established and emerging themes across all the five weeks (Figure 2B). Uniquely, Copilot had a strong thematic focus on themes closely associated with specific demographic groups, such as language barriers, disabilities, LGBTQ+, and aging populations. This suggests that Copilot excels in recognizing underserved populations within the broader context of healthcare disparities. Despite the higher consistency in theme recognition, Copilot's greater propensity for transitions between categorizing a given theme as established or emerging indicates an indecisive comprehension of the specific research field (Figure 2C).

Gemini behaves as a balanced approach between ChatGPT and Copilot, with fewer transitions and considerable consistency in identification of research themes such as environmental health, social media, and climate change, offering an integrative and interdisciplinary perspective on healthcare disparities.

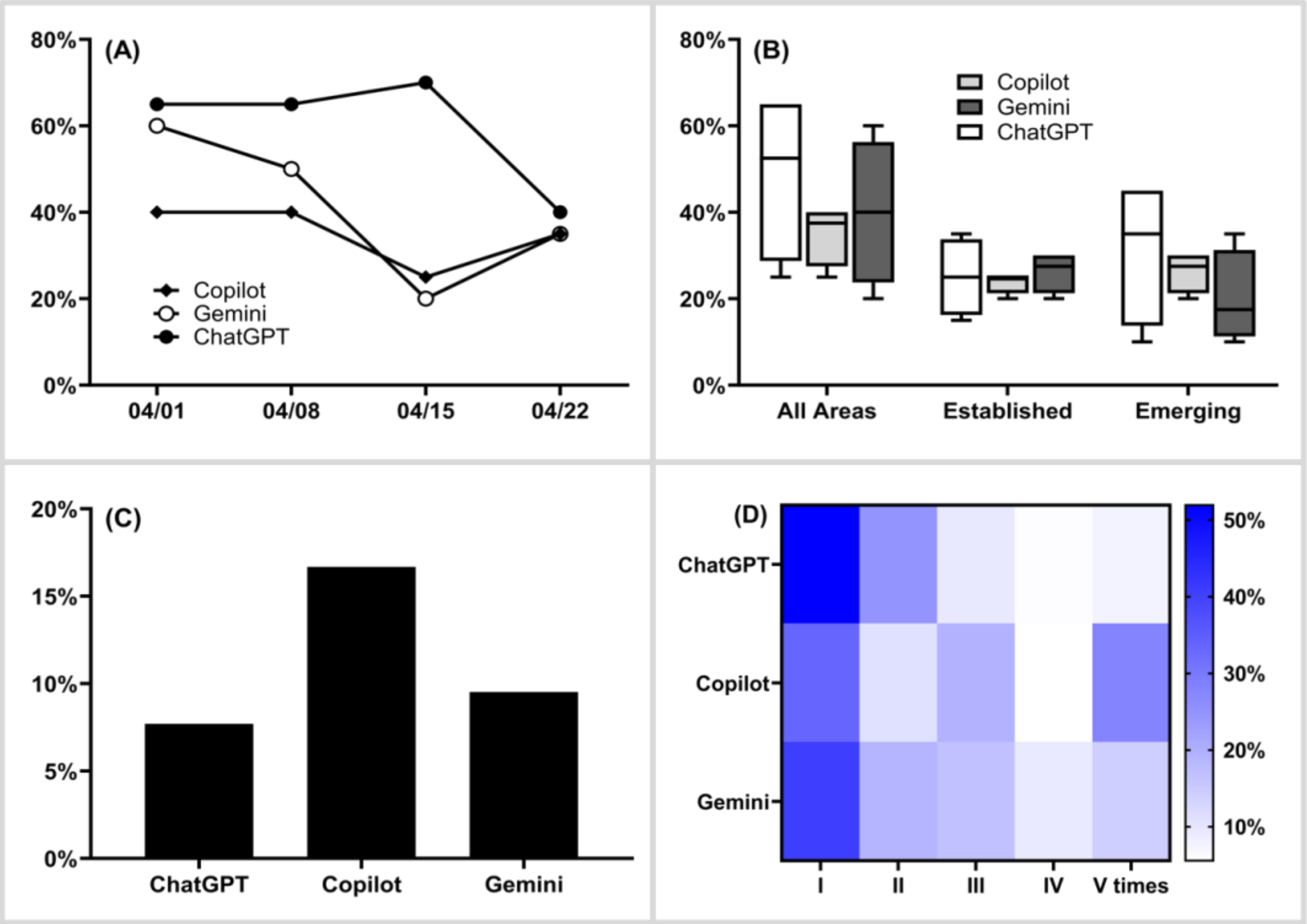

**Figure 2.** The temporal variation of LLM responses to research themes of healthcare disparities. (A) The trend line shows the percentage of themes that have been updated as compared to the previous week. (B) The bar chart shows the percentage changes compared to the previous responses in established and emerging themes throughout five weeks. (C) The column shows the percentage of themes that have transitioned between established and emerging status from week to week. (D) The heat map shows the aggregate proportion of themes which were recurred a certain number of times throughout the queries (i.e., if a theme was present in just 1 query or 5 queries - 1, 2, 3, 4 or 5 times).

Taking all these factors into consideration, it seems that Gemini is best suited for a flexible yet consistent identification of both well-established and emerging themes. Its moderate sensitivity to week-to-week changes and lower rate of transitions between established and emerging classifications make it reliable for maintaining a foundational understanding of critical disparities. In comparison with ChatGPT, which identified more than half of the topics only once or twice throughout the five weeks of queries (Figure 2D), Gemini reflected a more stable and balanced approach, generating key research themes and emerging areas consistently, offering a broad and reliable perspective on healthcare disparities.

Statistical analysis using the Kruskal-Wallis H test has supported this observation. It revealed no significant differences in H-index distributions across the healthcare disparities research themes identified by the three LLMs ($p$ = 0.6649). In practical terms, this suggests that the three models tend to identify research themes with comparable levels of scholarly influence, as measured by the H-index. Similarly, no significant differences were found in the classification of research themes as established or emerging ($p$ = 0.3461). This indicates that their ability to discern the maturity level of research themes is statistically comparable. However, the frequency of theme identification returned a $p$-value of 0.0515, marginally above the conventional alpha (0.05) threshold. While not statistically significant, this borderline result suggests a potential trend that might warrant further exploration with longer period or adjusted criteria. Collectively, the three LLMs show statistically comparable performance in identifying and classifying healthcare disparity research themes, with Gemini offering a more balanced and stable approach.

## Conclusions

This study reveals the evolving nature of LLMs in their generative responses to the time-series queries of research themes in healthcare disparities. The LLMs corroborated with the metric measure of H-index present their capacity to illustrate the state of healthcare inequities, offering insights into both persistent challenges and new frontiers in the field. The LLMs' consistent recognition of emerging themes with low H-indices holds promise for staying ahead of the research curve by focusing on areas poised for future growth. Yet, the variability in classification of themes identified only once might necessitate continuous updates and improvements to the LLM's training data or algorithms to enhance consistency and accuracy.

Comparison of the LLM outcomes highlights the characterization of the scope presented by each model. ChatGPT demonstrates high adaptability to emerging trends, while Copilot excels in consistency and stable recognition of research themes, and Gemini strikes a balance between the other two LLMs.

The practical impact of this study lies in leveraging LLMs' distinct strengths to facilitate the public and scholars for a comprehensive and efficient understanding of the ongoing research in healthcare disparities. Through monitoring and incorporating perspectives from different LLMs, attention can be brought to divergent emphases, gaps, and potential blind spots in the current body of research, thereby enabling a richer comprehension of the models' adaptive behaviour and allowing identification of emerging priorities to guide future studies more effectively.

Meanwhile, it acknowledges limitations that warrant future consideration when navigating a complex research landscape with LLM lens. Firstly, the relatively short period of study may not capture the longer-term changes regarding generalizability of the observed evolutionary trends. Secondly, the inclusion of only popular models may restrict generalizability of the findings across the broader spectrum of rapidly developing LLMs. Different models, particularly those with specialized training datasets or architectures, may carry varied capabilities in identifying and classifying research themes. Given these limitations, future studies could focus on the consistent follow-up with broader spectrum of models to fully encompass the entire dynamic landscape of healthcare disparities research.

## Acknowledgements

The author sincere thanks the editor and reviewers for their meticulous review, insightful comments, and constructive suggestions, which have significantly improved the quality of this manuscript.

## About the author

**David An** is an undergraduate at Harvard College and currently working as a researcher at Harvard Medical School pursuing his passion for biochemistry and molecular biology, with the goal of advancing scientific knowledge and improving human health.

## References

Abràmoff, M. D., Tarver, M. E., Loyo-Berrios, N., Trujillo, S., Char, D., Obermeyer, Z., & Eydelman, M. B., Foundational Principles of Ophthalmic Imaging and Algorithmic Interpretation Working Group of the Collaborative Community for Ophthalmic Imaging Foundation, W., DC, and Maisel, W. H. (2023). Considerations for addressing bias in artificial intelligence for health equity. *NPJ Digital Medicine*, 6(1), 170. https://www.nature.com/articles/s41746-023-00913-9

Aldoseri, A., Al-Khalifa, K. N., & Hamouda, A. M. (2023). Re-thinking data strategy and integration for artificial intelligence: concepts, opportunities, and challenges. *Applied Sciences*, 13(12), 7082. https://doi.org/10.3390/app13127082

Amirova, A., Fteropoulli, T., Ahmed, N., Cowie, M. R., & Leibo, J. Z. (2024). Framework-based qualitative analysis of free responses of large language models: algorithmic fidelity. *PloS One*, 19(3), e0300024. https://doi.org/10.1371/journal.pone.0300024

Argyle, L. P., Busby, E. C., Fulda, N., Gubler, J. R., Rytting, C., & Wingate, D. (2023). Out of one, many: using language models to simulate human samples. *Political Analysis*, 31(3), 337-351. https://doi.org/10.1017/pan.2023.2

Bailey, R., Sharpe, D., Kwiatkowski, T., Watson, S., Dexter Samuels, A., & Hall, J. (2018). Mental health care disparities now and in the future. Journal of Racial and Ethnic Health Disparities, 5, 351-356. https://doi.org/10.1007/s40615-017-0377-6

Beheshti, M., Toubal, I. E., Alaboud, K., Almalaysha, M., Ogundele, O. B., Turabieh, H., Abdalnabi, N., Boren, S. A., Scott, G. J., & Dahu, B. M. (2025). Evaluating the reliability of ChatGPT for health-related questions: a systematic review. *Informatics*, 12(1), 9. https://doi.org/10.3390/informatics12010009

Berkman, N. D., Davis, T. C., & McCormack, L. (2010). Health literacy: what is it? *Journal of Health Communication*, 15(S2), 9-19. https://doi.org/10.1080/10810730.2010.499985

Braveman, P. (2006). Health disparities and health equity: concepts and measurement. *Annual Review of Public Health*, 27, 167-194. https://doi.org/10.1146/annurev.publhealth.27.021405.102103

Brondolo, E., Gallo, L. C., & Myers, H. F. (2009). Race, racism and health: disparities, mechanisms, and interventions. *Journal of Behavioral Medicine*, 32, 1-8. https://doi.org/10.1007/s10865-008-9190-3

Busch, F., Hoffmann, L., Rueger, C., van Dijk, E. H. C., Kader, R., Ortiz-Prado, E., Makowski, M. R., Saba, L., Hadamitzky, M., Kather, J. N., Truhn, D., Cuocolo, R., Adams, L. C., & Bressem, K. K. (2025). Current applications and challenges in large language models for patient care: a systematic review. *Communications Medicine*, 5, 26. https://doi.org/10.1038/s43856-024-00717-2

Chang, Y., Wang, X., Wang, J., Wu, Y., Yang, L., Zhu, K., Chen, H., Yi, X., Wang, C., Wang, Y., & Ye, W. (2024). A survey on evaluation of large language models. *ACM Transactions on Intelligent Systems and Technology*, 15(3), 1-45. https://doi.org/10.1145/3641289

Clusmann, J., Kolbinger, F. R., Muti, H. S., Carrero, Z. I., Eckardt, J. N., Laleh, N. G., Löffler, C. M. L., Schwarzkopf, S. C., Unger, M., Veldhuizen, G. P., & Wagner, S. J. (2023). The future landscape of large language models in medicine. *Communications Medicine*, 3(1), 141. https://doi.org/10.1038/s43856-023-00370-1

Cook, B. L., Hou, S. S. Y., Lee-Tauler, S. Y., Progovac, A. M., Samson, F., & Sanchez, M. J. (2019). A review of mental health and mental health care disparities research: 2011-2014. *Medical Care Research and Review*, 76(6), 683-710. https://doi.org/10.1177/1077558718780592

Cook, B. L., McGuire, T., & Miranda, J. (2007). Measuring trends in mental health care disparities, 2000–2004. *Psychiatric Services*, 58(12), 1533-1540. https://doi.org/10.1176/ps.2007.58.12.1533

Costas, R., & Bordons, M. (2007). The H-index: advantages, limitations and its relation with other bibliometric indicators at the micro level. *Journal of Informetrics*, 1(3), 193-203. https://doi.org/10.1016/j.joi.2007.02.001

Fiscella, K., & Sanders, M. R. (2016). Racial and ethnic disparities in the quality of health care. *Annual Review of Public Health*, 37(1), 375-394. https://doi.org/10.1146/annurev-publhealth-032315-021439

FitzGerald, C., & Hurst, S. (2017). Implicit bias in healthcare professionals: a systematic review. *BMC Medical Ethics*, 18, 1-18. https://doi.org/10.1186/s12910-017-0179-8

Fullman, N., Yearwood, J., Abay, S. M., Abbafati, C., Abd-Allah, F., Abdela, J., Abdelalim, A., Abebe, Z., Abebo, T. A., Aboyans, V., & Abraha, H. N. (2018). Measuring performance on the Healthcare Access and Quality Index for 195 countries and territories and selected subnational locations: a systematic analysis from the Global Burden of Disease Study 2016. *The Lancet*, 391(10136), 2236-2271. https://doi.org/10.1016/S0140-6736(18)30994-2

Google. (2023). Bard updates from Google I/O 2023: Images, new features. Retrieved 15 June 2024, from https://blog.google/technology/ai/google-bard-updates-io-2023/ (Archived by the Internet Archive at https://web.archive.org/web/20240607074554/https://blog.google/technology/ai/google-bard-updates-io-2023/)

Kirubarajan, A., Patel, P., Leung, S., Prethipan, T., & Sierra, S. (2021). Barriers to fertility care for racial/ethnic minority groups: a qualitative systematic review. *F&S Reviews*, 2(2), 150-159. https://doi.org/10.1016/j.xfnr.2021.01.001

Maity, S., & Saikia, M. J. (2025). Large language models in healthcare and medical applications: a review. *Bioengineering*, 12, 631. https://doi.org/10.3390/bioengineering12060631

Meduri, K., Gonaygunta, H., Nadella, G. S., Pawar, P. P., & Kumar, D. (2024). Adaptive intelligence: GPT-powered language models for dynamic responses to emerging healthcare challenges. *International Journal of Advanced Research in Computer and Communication Engineering*, 13(1), 104-109. https://doi.org/10.17148/IJARCCE.2024.13114

Mehdi, Y. (2023). Announcing Microsoft Copilot, your everyday AI companion. Retrieved 15 June 2024, from https://blogs.microsoft.com/blog/2023/09/21/announcing-microsoft-copilot-your-everyday-ai-companion/ (Archived by the Internet Archive at https://web.archive.org/web/20240713133743/https://blogs.microsoft.com/blog/2023/09/21/announcing-microsoft-copilot-your-everyday-ai-companion/)

Miranda, J., McGuire, T. G., Williams, D. R., & Wang, P. (2008). Mental health in the context of health disparities. *American Journal of Psychiatry*, 165(9), 1102-1108. https://doi.org/10.1176/appi.ajp.2008.08030333

Nazer, L. H., Zatarah, R., Waldrip, S., Ke, J. X. C., Moukheiber, M., Khanna, A. K., Hicklen, R. S., Moukheiber, L., Moukheiber, D., Ma, H., & Mathur, P. (2023). Bias in artificial intelligence algorithms and recommendations for mitigation. *PLOS Digital Health*, *2*(6), e0000278. https://doi.org/10.1371/journal.pdig.0000278

Nazi, Z. A., & Peng, W. (2024). Large language models in healthcare and medical domain: A review. *Informatics*, *11*(3), 57. https://doi.org/10.3390/informatics11030057

Nogueira, L., White, K. E., Bell, B., Alegria, K. E., Bennett, G., Edmondson, D., Epel, E., Holman, E. A., Kronish, I. M., & Thayer, J. (2022). The role of behavioral medicine in addressing climate change-related health inequities. *Translational Behavioral Medicine*, *12*(4), 526-534. https://doi.org/10.1093/tbm/ibac005

Norris, M., & Oppenheim, C. (2010). The H-index: a broad review of a new bibliometric indicator. *Journal of Documentation*, 66(5), 681-705. https://doi.org/10.1108/00220411011066790

Ong, J. C. L., Seng, B. J. J., Law, J. Z. F., Low, L. L., Kwa, A. L. H., Giacomini, K. M., & Ting, D. S. W. (2024). Artificial intelligence, ChatGPT, and other large language models for social determinants of health: current state and future directions. *Cell Reports Medicine*, 5(1), 101356. https://doi.org/10.1016/j.xcrm.2023.101356

OpenAI. (2024). ChatGPT4.0 [Large language model]. Retrieved 15 June 2024, from https://openai.com/chatgpt/ (Archived by the Internet Archive at https://web.archive.org/web/20240615082924/https://openai.com/chatgpt/)

Ray, P. P. (2023). ChatGPT: A comprehensive review on background, applications, key challenges, bias, ethics, limitations and future scope. *Internet of Things and Cyber-Physical Systems*, 3, 121-154. https://doi.org/10.1016/j.iotcps.2023.04.003

Riley, W. J. (2012). Health disparities: gaps in access, quality and affordability of medical care. *Transactions of the American Clinical and Climatological Association*, 123, 167. https://pmc.ncbi.nlm.nih.gov/articles/PMC3540621/ (Archived by the Internet Archive at https://web.archive.org/web/20240816175456/https://pmc.ncbi.nlm.nih.gov/articles/PMC3540621/)

Ruprecht, M. M., Wang, X., Johnson, A. K., Xu, J., Felt, D., Ihenacho, S., Stonehouse, P., Curry, C. W., DeBroux, C., Costa, D., & Phillips Ii, G. (2021). Evidence of social and structural COVID-19 disparities by sexual orientation, gender identity, and race/ethnicity in an urban environment. *Journal of Urban Health*, 98(1), 27-40. https://doi.org/10.1007/s11524-020-00497-9

Saeed, S. A., & Masters, R. M. (2021). Disparities in health care and the digital divide. *Current Psychiatry Reports*, 23, 1-6. https://doi.org/10.1007/s11920-021-01274-4

Schlicht, I. B., Zhao, Z., Sayin, B., Flek, L., & Rosso, P. (2025). Do LLMs provide consistent answers to health-related questions across languages? In European Conference on Information Retrieval (pp. 314-322). https://doi.org/10.1007/978-3-031-88714-7_30

Shah, F. A., & Jawaid, S. A. (2023). The H-index: an indicator of research and publication output. *Pakistan Journal of Medical Sciences*, 39(2), 315-316. https://doi.org/10.12669/pjms.39.2.7398

Singh, N., Lawrence, K., Richardson, S., & Mann, D. M. (2023). Centering health equity in large language model deployment. *PLOS Digital Health*, *2*(10), e0000367. https://doi.org/10.1371/journal.pdig.0000367

Singhal, K., Azizi, S., Tu, T., Mahdavi, S. S., Wei, J., Chung, H. W., Scales, N., Tanwani, A., Cole-Lewis, H., Pfohl, S., & Payne, P. (2023). Large language models encode clinical knowledge. *Nature*, *620*(7972), 172-180. https://doi.org/10.1038/s41586-023-06291-2

Thirunavukarasu, A. J., Ting, D. S. J., Elangovan, K., Gutierrez, L., Tan, T. F., & Ting, D. S. W. (2023). Large language models in medicine. *Nature Medicine*, 29(8), 1930-1940. https://doi.org/10.1038/s41591-023-02448-8

Veinot, T. C., Mitchell, H., & Ancker, J. S. (2018). Good intentions are not enough: how informatics interventions can worsen inequality. *Journal of the American Medical Informatics Association*, 25(8), 1080-1088. https://doi.org/10.1093/jamia/ocy052

Wang, D., & Zhang, S. (2024). Large language models in medical and healthcare fields: applications, advances, and challenges. *Artificial Intelligence Review*, 57(11), 299. https://doi.org/10.1007/s10462-024-10921-0

Wang, L., Wan, Z., Ni, C., Song, Q., Li, Y., Clayton, E., Malin, B., & Yin, Z. (2024). Applications and concerns of ChatGPT and other conversational large language models in health care: systematic review. *Journal of Medical Internet Research*, 26, e22769. https://doi.org/10.2196/22769

Yang, R., Tan, T. F., Lu, W., Thirunavukarasu, A. J., Ting, D. S. W., & Liu, N. (2023). Large language models in health care: Development, applications, and challenges. *Health Care Science*, 2(4), 255-263. https://doi.org/10.1002/hcs2.61

Yu, E., Chu, X., Zhang, W., Meng, X., Yang, Y., Ji, X., & Wu, C. (2025). Large language models in medicine: applications, challenges, and future directions. *International Journal of Medical Sciences*, 22(11), 2792-2801. https://doi.org/10.7150/ijms.111780